\documentclass[reprint,twocolumn,superscriptaddress,amsmath,amssymb,aps,prb]{revtex4-1}

\usepackage{graphicx}
\usepackage{graphics}
\usepackage{amsmath}
\usepackage{dcolumn}
\usepackage{bm}

\begin{document}


\title{Biaxial-stress driven tetragonal symmetry breaking in and high-temperature ferromagnetic semiconductor from half-metallic CrO$_2$}

\author{Xiang-Bo Xiao}
\affiliation{Beijing National Laboratory for Condensed Matter Physics, Institute of Physics, Chinese Academy of Sciences, Beijing 100190, China}
\affiliation{School of Physical Sciences, University of Chinese Academy of Sciences, Beijing 100190, China}
\author{Bang-Gui Liu}\email{bgliu@iphy.ac.cn}
\affiliation{Beijing National Laboratory for Condensed Matter Physics, Institute of Physics, Chinese Academy of Sciences, Beijing 100190, China}
\affiliation{School of Physical Sciences, University of Chinese Academy of Sciences, Beijing 100190, China}

\date{\today}

\begin{abstract}
It is highly desirable to combine the full spin polarization of carriers with modern semiconductor technology for spintronic applications. For this purpose, one needs good crystalline ferromagnetic (or ferrimagnetic) semiconductors with high Curie temperatures. Rutile CrO$_2$ is a half-metallic spintronic material with Curie temperature 394 K and can have nearly-full spin polarization at room temperature. Here, we find through first-principles investigation that when a biaxial compressive stress is applied on rutile CrO$_2$, the density of states  at the Fermi level decreases with the in-plane compressive strain, there is a structural phase transition to an orthorhombic phase at the strain of -5.6\%, and then appears an electronic phase transition to a semiconductor phase at -6.1\%. Further analysis shows that this structural transition, accompanying the tetragonal symmetry breaking, is induced by the stress-driven distortion and rotation of the oxygen octahedron of Cr, and the half-metal-semiconductor transition originates from the enhancement of the crystal field splitting due to the structural change. Importantly, our systematic total-energy comparison indicates the  ferromagnetic Curie temperature remains almost independent of the strain, near 400 K. This biaxial stress can be realized by applying biaxial pressure or growing the CrO$_2$ epitaxially on appropriate substrates. These results should be useful for realizing full (100\%) spin polarization of controllable carriers as one uses in modern semiconductor technology.
\end{abstract}

\pacs{Valid PACS appear here}
\maketitle


\section{Introduction}

It is always of much interest to search for high-performance spintronic materials\cite{r1,r2,r3,r4}. Full (100\%) spin polarization can be realized in half-metallic materials\cite{hm1,r4} such as rutile CrO$_2$\cite{co1,co2,co3,co4}, because one of the two spin channels is metallic and the other has a semiconductor gap. For rutile CrO$_2$, ferromagnetic Curie temperature can reach to 394 K and great effort has been made to understand the physics in it and achieve high-quality materials for full spin polarization\cite{co5,co6,co7,co8,co9,co10,co11,1998,2005}. It is also interesting to seek half-metallic materials based on semiconductors\cite{r4,sh1,sh2,sh3,shm1,shm2,shm3}. Furthermore, one naturally hopes to combine the full polarization with modern semiconductor technology. For this purpose, one needs crystalline ferromagnetic (or ferrimagnetic) semiconductors with high Curie temperatures and then chooses appropriate dopants for controllable carriers. In this direction, many good materials have been found, and among them, perovskite Ca$_2$CrReO$_6$, Ca$_2$FeReO$_6$, and Sr$_2$CrOsO$_6$ are ferrimagnetic semiconductors with high Curie temperatures 360 K, 522 K, and 725 K, respectively\cite{fs1,fs2,fs3,fs4}. Along this line, it is hopeful to realize full (or at least nearly full) spin polarization for controllable concentrations of high-mobility carriers beyond room temperature.

Here, we present strain-induced tetragonal symmetry breaking in rutile CrO$_2$ and high-temperature ferromagnetic semiconductor phase from the half-metallic CrO$_2$ in terms of first-principles calculation. Biaxial compressive stress is applied to rutile CrO$_2$ to make the in-plane lattice constants ($a$, $b$) reduce by up to -8\%, with the other lattice constant ($c$) determined by stress equilibrium. The stress can make the horizontal bonds (Cr-O and O-O) compress substantially and keep the side bond-angle (between two oblique bonds) decrease notably, but the oblique bonds (Cr-O) remain almost unchanged. With the stress increasing, the density of states at the Fermi level decreases substantially, and finally the rutile CrO$_2$ becomes an orthorhombic phase when $a$ is compressed by 5.6\%. Keeping the biaxial stress increasing, there will be an electronic phase transition to ferromagnetic semiconductor phase. It is very interesting that for all the compression ratios, the magnetic energies per formula unit remain almost unchanged and then the Curie transition temperature still remains near 400 K. Our analysis indicates that these phase transitions are driven by the distortion and rotation of the oxygen octahedron of Cr and the resulting crystal field splitting enhancement.
This could open a door to achieve crystalline ferromagnetic semiconductors with high Curie temperatures. More detailed results will be presented in the following.

\section{Computational details}

Our density-functional-theory calculations are done with the Vienna package WIEN2K \cite{wien2k}, which is a full-potential augmented plane wave plus local orbital program within the density functional theory \cite{dft1,dft2}. For total energy calculation of rutile and orthorhombic CrO$_2$, we take the generalized-gradient approximation (GGA) \cite{pbe} for the exchange-correlation potential. The scalar approximation is taken for the relativistic effects of valence states. For improving the electronic structure calculations, we use modified Becke-Johnson (mBJ) \cite{mbj} exchange potential plus GGA correlation potantial for the exchange-correlation potential. The muffin-tin radii ($R_{mt}$) of Cr and O atom are optimized to achieve high accuracy. We use 3000 k points in the Brillouin zone and set $R_{mt}\times K_{max}$ to 7.0, and in addition, we make angular momentum expansion up to $l_{max}$=10 in the muffin tins. When the integrated absolute value of the charge difference per formula unit, $\int |\rho_n -\rho_{n-1}|dr$, between the input charge density [$\rho_{n-1}(r)$] and the output charge density [$\rho_n(r)$] is less than $0.0001|e|$ where $e$ is the electron charge, the self-consistent calculation will be considered to be converged. The spin-orbit coupling is also taken into account to investigate the spin polarization at the Fermi level.

\begin{figure}[!htbp]
{\centering  
\includegraphics[clip, width=7cm]{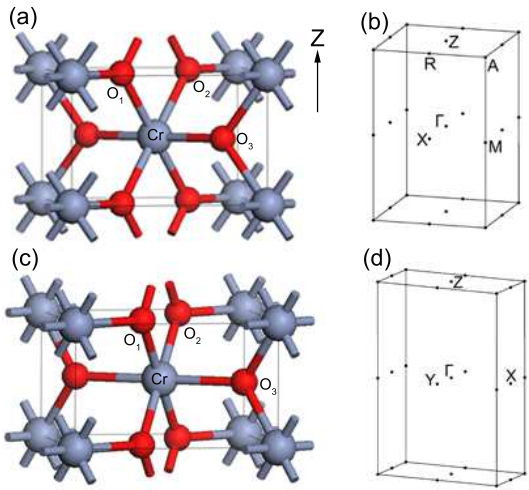}}\\
\caption{(Color online) The crystal structure (a) and the Brillouin zone (b) of rutile phase (P4$_2$/mnm), and the crystal structure (c) and the Brillouin zone (d) of the orthorhombic phase (Pnnm) of CrO$_2$.}\label{fig1}
\end{figure}

\section{Results and discussion}

\subsection{Strain-dependent structure parameters}

Rutile CrO$_2$ assumes a tetrahedral structure with the space group P4$_2$/mnm (\#136)\cite{co1,co2,co3,co4}, as shown in Fig. 1. There are two formula units in the unit cell. Two chromium atoms are located at (0,0,0) and ($\frac 12$,$\frac 12$,$\frac 12$) in the internal coordinates, and four oxygen atoms at ($u$,$u$,$0$), (1-$u$,1-$u$,0), ($\frac 12$+$u$,$\frac 12$-$u$,$\frac 12$), and ($\frac 12$-$u$,$\frac 12$+$u$,$\frac 12$), respectively, where $u$ ($\le 1$) is a dimensionless internal coordinate parameter.
We have fully optimized the lattice constants ($a$ and $c$) and the internal coordinate parameter $u$ by using GGA. The resulting equilibrium structural parameters ($a_0$, $c_0$, and $u_0$) are equivalent to 4.4507\AA, 2.9317\AA, and 0.304, respectively. The electronic structure is half-metallic, which is consistent with previous DFT calculated results and experimental observations \cite{co1,co2,co3,co4}.

Because large uniaxial stress can make the rutile phase transit to an orthorhombic phase (Pnnm, No. 58), we also plot its crystal structure and corresponding Brillouin zone in Fig. 1. For this structure under uniaxial stress, we have three lattice constants ($a$, $b$, and $c$) and two internal coordinate parameters ($u_x$,$u_y$). Because the stress is always uniaxial, it is convenient to describe the in-plane strain in a unified way. For this purpose, we introduce a parameter $d$ to equal $a$ for the rutile phase and $\sqrt{ab}$ for the orthorhombic phase, and then use $\Delta d/d_0=(d-d_0)/d_0$, with $d_0=a_0$, to express the in-plane strain.

\begin{table}[!h]
\caption{Optimized structural parameters ($a$, $b$, $c$, $u_x$, and $u_y$), total energy per formula unit ($E$), and semiconductor gap ($E_g$) of CrO$_2$ for different in-plane lattice constant $d$ and its relative change $\Delta d/d_0$ as defined in the text. The semiconductor gap ($E_g$) values are calculated with mBJ.}
\begin{ruledtabular}
\begin{tabular}{ccccccc}
$\Delta d/d_0$(\%) &  $a$ (\AA) & $b$ (\AA)  & $c$ (\AA) &  ($u_x$,$u_y$)  &$E$ (eV) &  $E_g$ (eV)\\ \hline
0     &  4.451 & -  & 2.932  &   (0.304,0.304)  & 0    &   - \\
-3.0  &  4.317 & -  & 3.000  &   (0.305,0.305)  & 0.130&   - \\
-5.0  &  4.228 & -  & 3.046  &   (0.306,0.306)  & 0.386&   - \\
-5.5  &  4.206 & -  & 3.062  &   (0.306,0.306)  & 0.474&   - \\
-5.8  &  4.214 & 4.172  & 3.061  &  (0.309,0.301)  & 0.530&   - \\
-6.0  &  4.226 & 4.143  & 3.065  &  (0.314,0.297)  & 0.566&   - \\
-6.5  &  4.234 & 4.091  & 3.075  &  (0.319,0.292)  & 0.676&   0\\
-7.0  &  4.221 & 4.059  & 3.087  &  (0.322,0.290)  & 0.786&   0.03\\
-8.0  &  4.216 & 3.977  & 3.109  &  (0.328,0.284)  & 1.044&   0.05
\end{tabular}
\end{ruledtabular}
\end{table}

In order to manipulate the electronic structure, we can apply a biaxial stress to the rutile CrO$_2$. Such stress can be realized by applying pressure along the x and y axes or growing the rutile CrO$_2$ as epitaxial [001] thin films on some suitable substrates. Since tensile stress has limited effect on the electronic states of the CrO$_2$ near the Fermi level, we will concentrate on compressive stress within the x-y plane. It is clear that $\Delta d/d_0=\Delta a/a_0=0$ corresponds to the equilibrium lattice. For the rutile phase, with a given in-plane lattice constant $a<a_0$, we make a full structural optimization to determine the other lattice constant $c$ and the internal parameter $u$. For the orthorhombic phase, with $d=\sqrt{ab}$ used to correspond the $a$ parameter in the rutile phase, we also make full optimization to determine the five parameters: $a$, $b$, $c$, $u_x$, and $u_y$. The total energy of the strained CrO$_2$ is calculated with the optimized strained crystal structure. We have done a series of optimization with $0>\Delta d/d_0\ge -8$\%. The structural parameters ($a$, $b$, $c$, $u_x$, and $u_y$) and total energy ($E$) values are summarized in Table I.

\begin{figure}[!htbp]
\centering  
\includegraphics[clip, width=8cm]{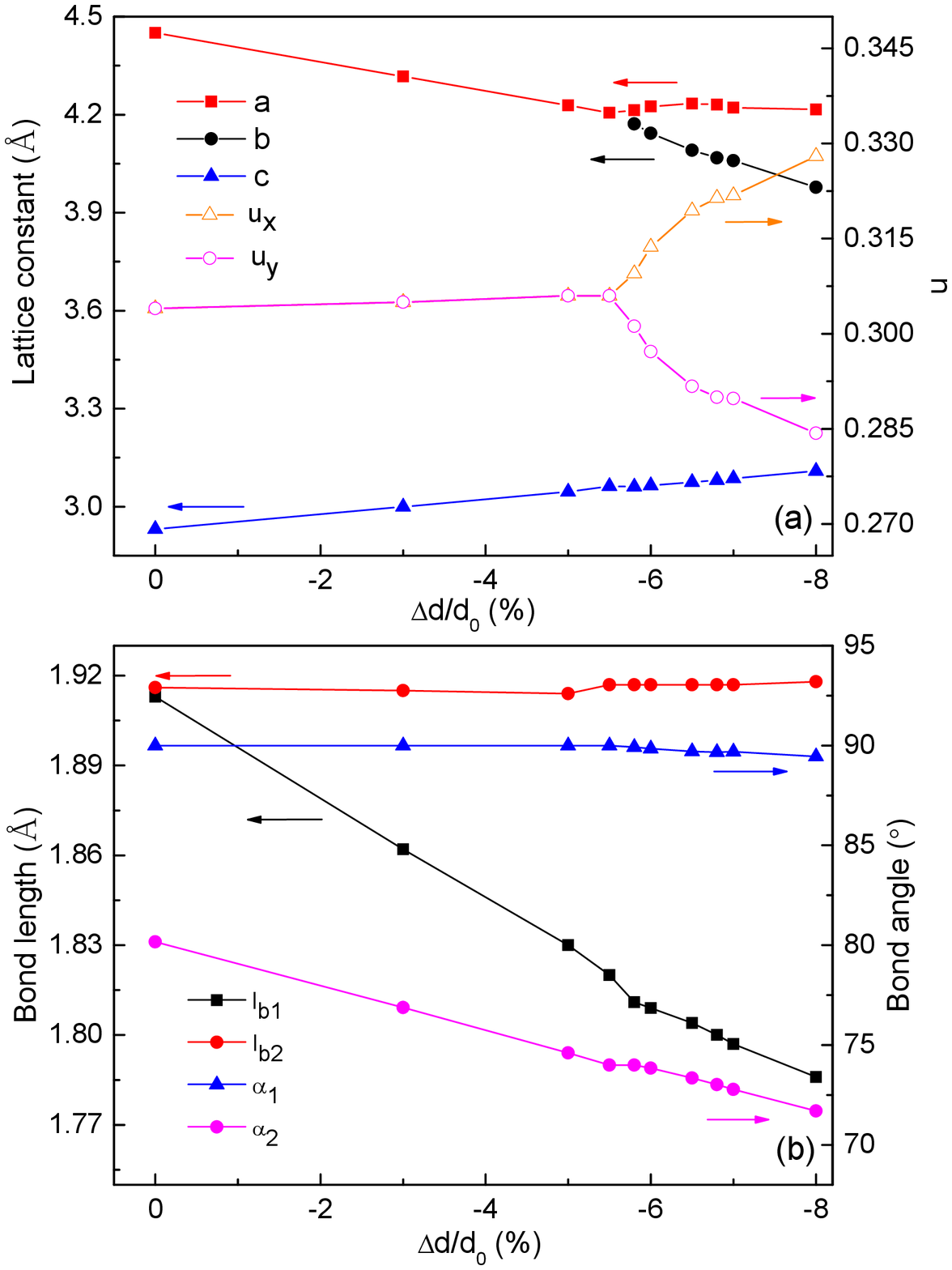}\\
\caption{(Color online) The $\Delta d/d_0$ dependence of the structural parameters ($a$, $b$, $c$, $u_x$, $u_y$) (a) and the bond parameters ($l_{b1}$, $l_{b2}$, $\alpha_1$, $\alpha_2$) (b). The in-plane lattice constants and $u$ parameters are different along the x and y axes in the orthorhombic phase, and the structural phase transition can also be seen in the curves of bond lengths and angles. }\label{fig2}
\end{figure}

At first, we assume that CrO$_2$ is in the rutile phase. As the in-plane compressive strain ($-\Delta d/d_0$) increases ($d=a$ decreases), $c$ increases, and the internal parameter $u$ increases slightly. The opposite trend of $a$ and $c$ reflects the trend that with stress applied, the volume tries to remain unchanged. Because $c$ becomes large with the biaxial compression increasing, there is actually a tensile strain in the z axis. The stronger the compression is, the larger the energy $E$ becomes, which is expected because the strain drive the CrO$_2$ to deviate from the equilibrium structure. When the in-plane strain reaches to -5.6\%, the rutile phase will transit to the orthorhombic phase, keeping the half-metallic feature.
More importantly, there is a phase transition from the half-metallic phase to a semiconductor when the compression ratio of $d$ reaches to -6.1\%. We visualize these trends of $a$, $b$, $c$, $u=(u_x,u_y)$ in Fig. 2a by plotting them as functions of the compression ratio, $\Delta d/d_0$. It is clear that the lattice constants change almost linearly with $\Delta d/d_0$ decreasing and the parameter $u$ has little change until $\Delta d/d_0$ reaches to the structural phase transition point. Entering the orthorhombic phase, both $a-b$ and $u_x-u_y$ increase notably.

\begin{table}[!h]
\caption{The bond angles ($\alpha_1$ and $\alpha_2$) and bond lengths ($l_{b1}$ and $l_{b2}$) for  different $\Delta d/d_0$ values. }
\begin{ruledtabular}
\begin{tabular}{ccccc}
$\Delta d/d_0$ (\%)  &  $\alpha_2$ ($^\circ$) &  $\alpha_1$ ($^\circ$) &  $l_{b1}$(\AA) &  $l_{b2}$(\AA)  \\ \hline
0     &    80.17   &90.00  &   1.913        &  1.916       \\
-3.0  &    76.88   &90.00  &   1.862        &  1.915       \\
-5.0  &    74.60   &90.00  &   1.830        &  1.914      \\
-5.5  &    74.00   &90.00  &   1.820        &  1.917      \\
-5.8  &    73.98   &89.92  &   1.811        &  1.917    \\
-6.0  &    73.84   &89.85  &   1.809        &  1.917     \\
-6.5  &    73.35   &89.71  &   1.804        &  1.917      \\
-7.0  &    72.77   &89.69  &   1.797        &  1.917      \\
-8.0  &    71.70   &89.45  &   1.786        &  1.918
\end{tabular}
\end{ruledtabular}
\end{table}

It is vital to check the stability of ferromagnetic order and Curie temperature against the biaxial compression. Because the biaxial compression enhances the ferromagnetic exchange in the x-y plane and maybe weakens the ferromagnetic exchange in the z direction, we have constructed antiferromagnetic order along the z direction and compared their total energies with the ferromagnetic rutile or orthorhombic CrO$_2$ for $\Delta d/d_0=$0, -3.0, -5.0, -6.0, -7.0, -8.0\%, respectively. For the  $\Delta d/d_0$ values, our calculated results show that the ferromagnetic CrO$_2$ is lower than the antiferromagnetic configuration by 0.320, 0.346, 0.352, 0.369, 0.373, and 0.378 eV per formula unit, respectively. Therefore, the ferromagnetic order is stable against spin fluctuation, and when the biaxial compression is applied, the Curie temperature should remain approximately the same as that of the equilibrium structure \cite{r4,co9,co10,co11}.

In order to find out the relationship between the electronic properties and the crystal structure, we also present in Table II the bond angles ($\alpha_1$ and $\alpha_2$) and the bond lengths ($\l_{b1}$ and $\l_{b2}$) of CrO$_2$ for different $\Delta d/d_0$ values. Here, $\alpha_1$ and $\alpha_2$ denote the O1-Cr-O3 and O1-Cr-O2 bond angles, and $\l_{b1}$ and $\l_{b2}$ indicate the lengths of the Cr-O3 and Cr-O2 bonds, respectively. It should be noted that the Cr-O3 bond is always perpendicular to the Cr-O1 and Cr-O2 bonds for the rutile phase, and the angle deviates from 90$^\circ$ for the orthorhombic phase. We plot in Fig. 2b the bond lengths and the bond angles as functions of different compression ratio of lattice constant, $\Delta d/d_0$. It is clear that the bond angle $\alpha_2$ and the bond length $\l_{b1}$ almost linearly decrease with $\Delta d/d_0$ changing from 0 to -8\%, whereas the other bond angle $\alpha_1$ and bond length $\l_{b2}$ changes only a little. Our analysis shows that the almost-linear decrease of the Cr-O3 bond length is caused directly by the decrease of the lattice constant $d$, and the Cr-O2 bond length remains almost unchanged because the decrease of the bond angle $\alpha_2$ compensates the opposite changes of $d$ and $c$.

\begin{figure}[!htbp]
\centering  
\includegraphics[clip, width=7.5cm]{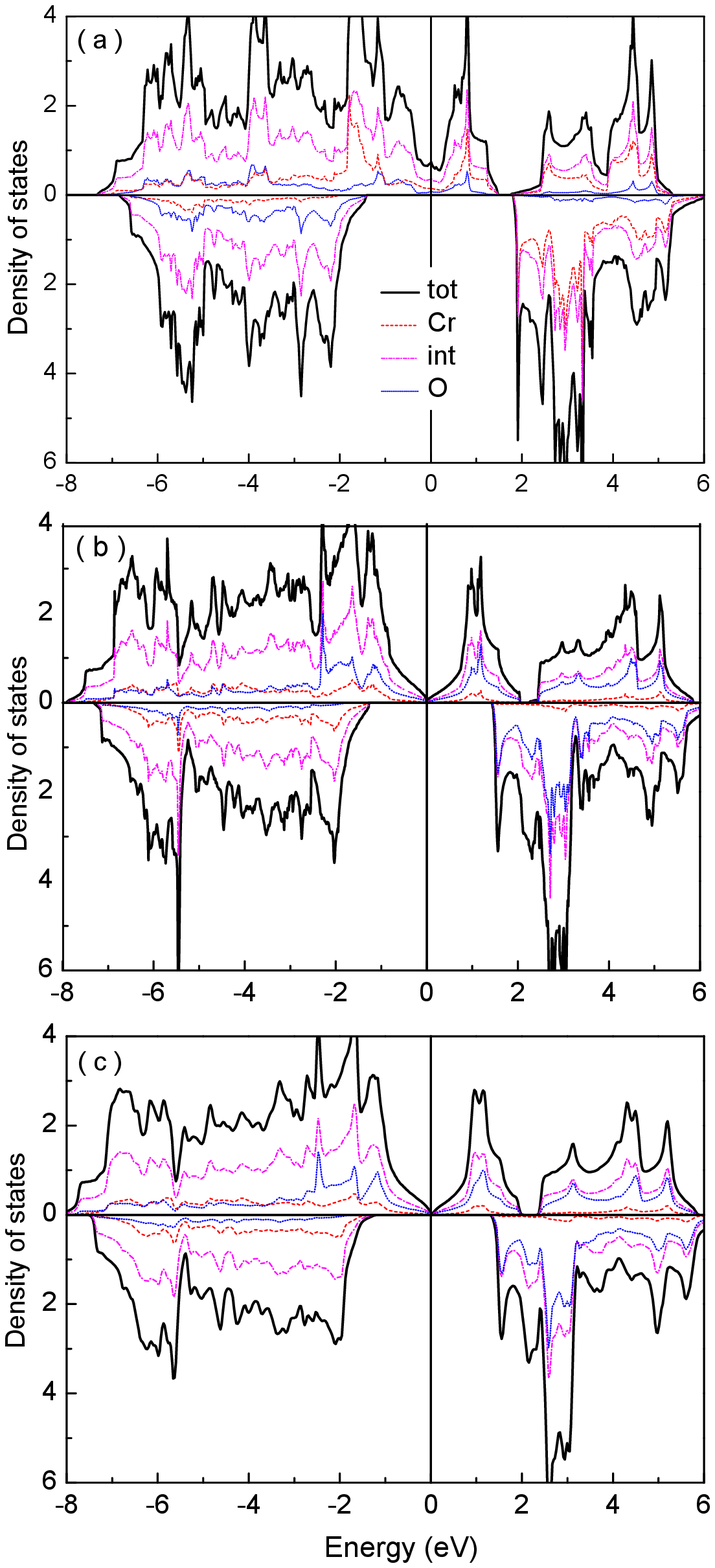}\\
\caption{(Color online) Spin-resolved total and partial density of states of CrO$_2$ for (a) the rutile equilibrium structure, (b) the orthorhombic structure with $d$ compressed by 6\%, and (c) the orthorhombic structure with $d$ compressed by 8\%.}\label{edge}
\end{figure}

\subsection{Electronic structures and phase transition}

We present in Fig. 3 the total and partial density of states (DOS) calculated with mBJ for three $\Delta d/d_0$ values: 0, -6.0\%, and -8.0\%. For the equilibrium structure (Fig. 3a), we found through our mBJ calculations that there is a big gap in the minority spin channel around the Fermi level, but the band structure is metallic in the majority spin channel, which is consistent with the fact that rutile CrO$_2$ is half-metallic \cite{co1,co2,co3,co4}. As we decrease the lattice parameter $d$ (or $a$), the total density of states at the Fermi level reduces in the majority spin channel, and we will meet the structural phase transition point and then obtain the orthorhombic phase. Fig. 3b shows the density of states for $\Delta d/d_0=-6.0$\%. When the lattice parameter $d$ is compressed by more than 6.1\%, it can be seen that the total density of states at the Fermi level becomes zero, as shown in Fig. 3c for $\Delta d/d_0=-8.0$\%, and consequently we obtain a ferromagnetic semiconductor phase. These electronic structures show an electronic phase transition from half-metal of rutile CrO$_2$ to ferromagentic semiconductor of the orthorhombic CrO$_2$ material. The low DOS near the Fermi level reflects that the O $p$ electrons are distributed mainly out of the O muffin tin, which is consistent with the large DOS contribution from the interstitial region.

\begin{figure}[!htbp]
\centering  
\includegraphics[clip, width=8cm]{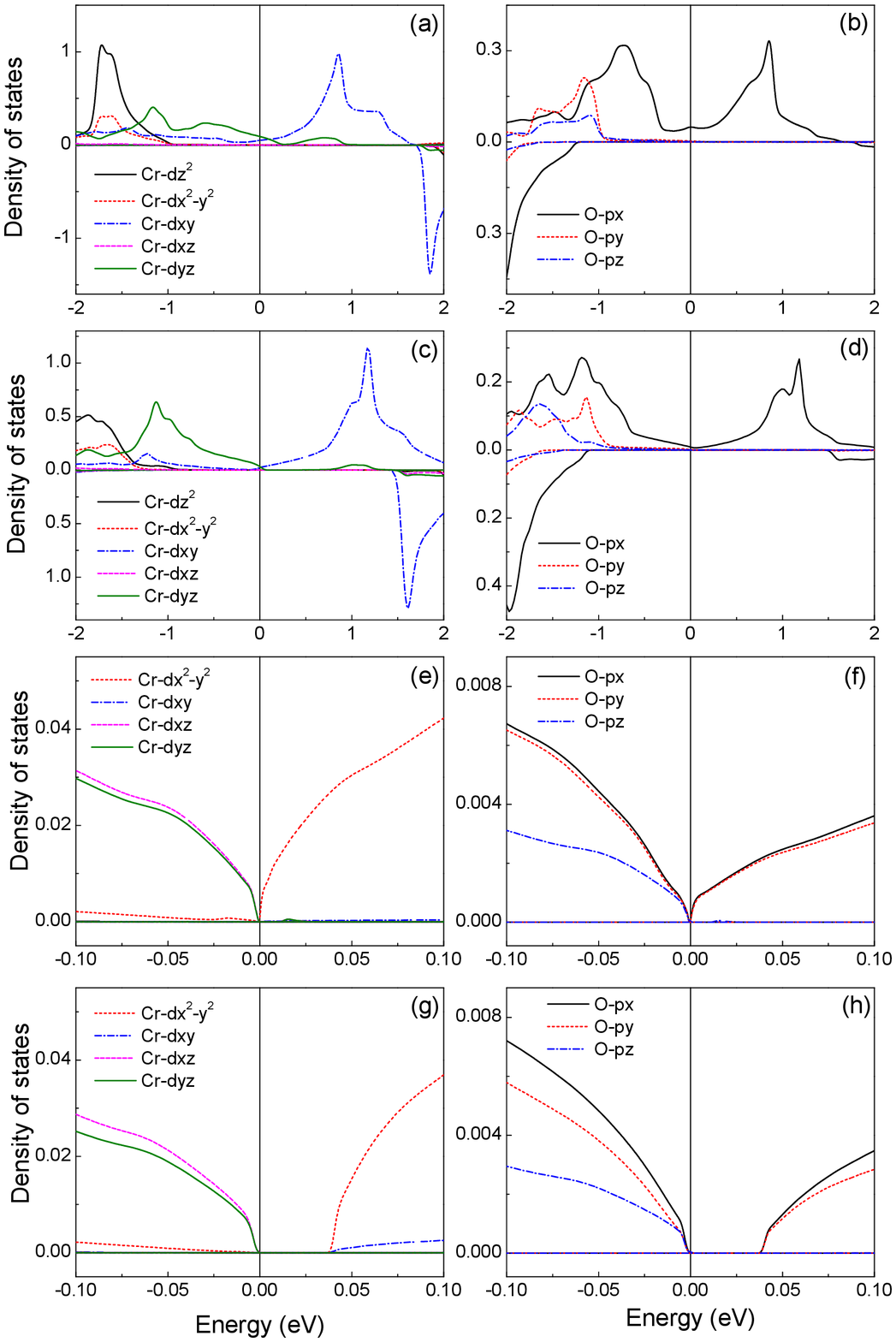}\\
\caption{(Color online) Orbital-resolved partial density of states of Cr-$d$ and O-$p$ (a,b) for the rutile equilibrium structure, (c,d) for the rutile phase with $d$ compressed by 5\%, (e,f) for the orthorhombic structure with $d$ compressed by 6\%, and (g,h) for the orthorhombic structure with $d$ compressed by 8\%.}\label{edge}
\end{figure}

In order to investigate the electronic properties of these systems in detail, we have also calculated the orbit-resolved density of states. The detailed orbit-resolved density of states in the vicinity of the Fermi level for $\Delta d/d_0=$0, -5.0\%, -6.0\%, and -8\% are shown in Fig. 4. Here, we use the usual local coordinates and orbital convention for rutile and orthorhombic structures\cite{1998,2005}. For the equilibrium structure, the Cr 3$d$ states near the Fermi level consist of Cr $d_{xy}$ and $d_{yz}$ in the majority spin channel, as shown in Fig. 4a. As for the O 2$p$ states, the $p_x$ orbital is dominant near the Fermi level, as shown in Fig. 4b. As the lattice parameter $d$ decreases, the main weights of the Cr $d_{xy}$ and $d_{yz}$ and O 2$p$ orbitals gradually move away from the Fermi level and the density of states there becomes smaller and smaller. The results for $\Delta d/d_0=-5.0$\% are presented in Figs. 4c and 4d. At $\Delta d/d_0=-6.0$\%, as shown in Figs. 4e and 4f, the density of states at the Fermi level is nearly zero, and the orbitals there are $d_{x^2-y^2}$, $d_{xz}$, $d_{yz}$, and the three $p$ orbitals. At $\Delta d/d_0=-6.1$\%, there is a gap open at the Fermi level, and the gap grows when the $d$ is compressed further, reaching to 0.05 eV at $\Delta d/d_0=-8.0$\%, as shown in Figs. 4g and 4h. These imply that when $\Delta d/d_0$ is beyond -6.1\%, the valence band top is occupied mainly by Cr $d_{xz}$ and $d_{yz}$ orbitals, and the conduction band bottom mainly by Cr $d_{x^2-y^2}$.

\begin{figure}[!htbp]
\centering  
\includegraphics[clip, width=8.6cm]{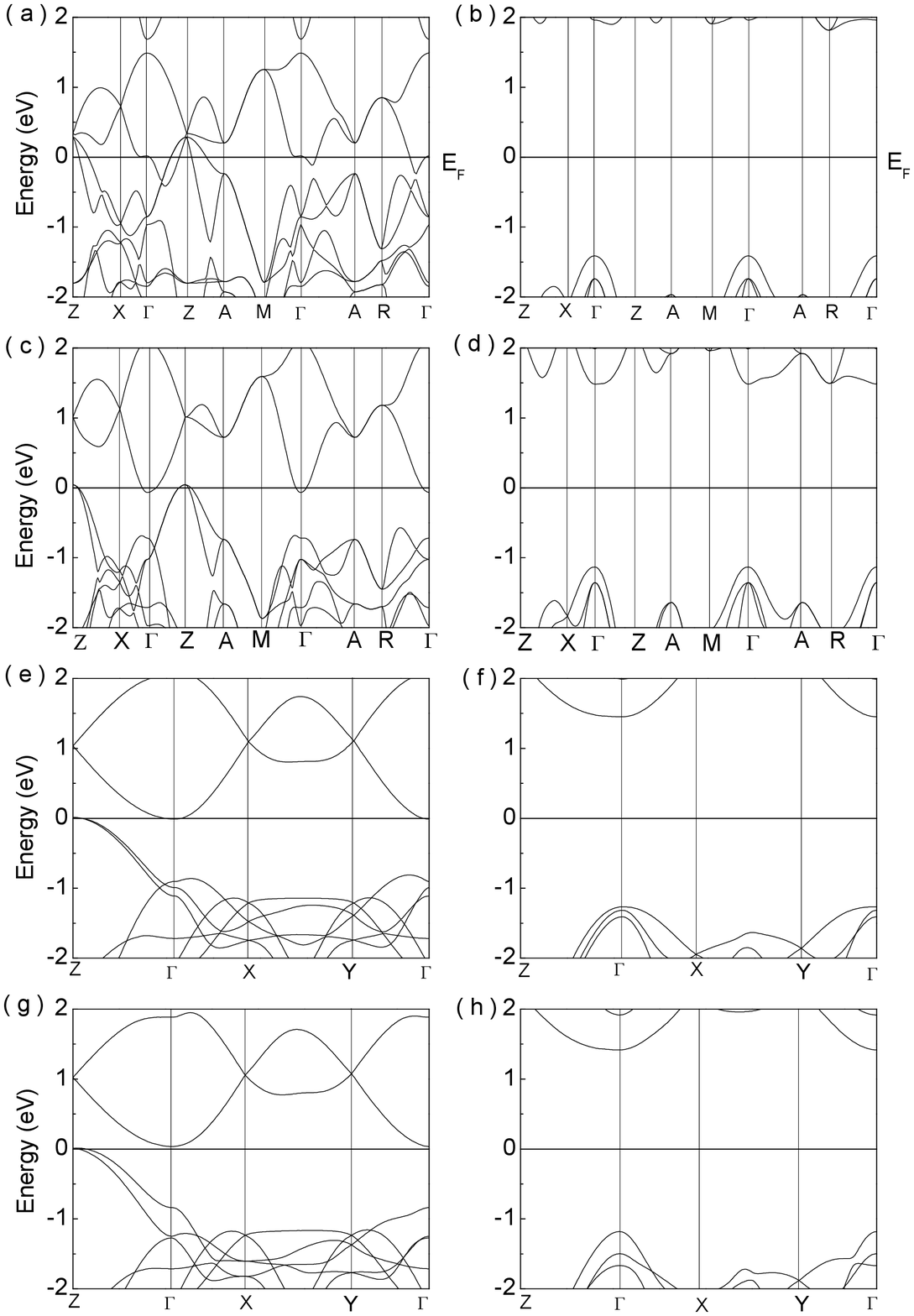}\\
\caption{Spin-polarized band structures of CrO$_2$ (a,b) for the rutile equilibrium structure, (c,d) for the rutile phase with $d$ compressed by 5\%, (e,f) for the orthorhombic structure with $d$ compressed by 6\%, and (g,h) for the orthorhombic structure with $d$ compressed by 8\%. The left column presents the majority-spin band structure, and the right column the minority-spin one.}\label{edge}
\end{figure}

We present in Fig. 5 the band structures of strained CrO$_2$ for $\Delta d/d_0=$0, -5.0\%, -6.0\%, and -8\%. It is clear that without any strain, the rutile CrO$_2$ is a typical half-metal with a large gap in the minority-spin channel, as shown in Figs. 5a and 5b. As we decrease the lattice parameter $d$, the bands near the Fermi level tend to move away from the Fermi level in the majority-spin channel, with the gap still remaining  wide in the minority-spin channel. The majority-spin gap is approximately at the middle of the minority-spin gap. With $\Delta d/d_0=$-5\%$\sim$-6\%, the conduction band minimum is at the $\Gamma$ point in the Brillouin zone, and the valence band maximum at the Z point, which implies that the strained CrO$_2$ is a semi-metal, as shown in Figs. 5(c-f). The transition happens at $\Delta d/d_0=-6.1$\%, from which on there is a narrow gap open at the Fermi level in the majority-spin channel, and then the strained CrO$_2$ is a semiconductor with an indirect gap. Combining with the DOS near the Fermi level, we can conclude that for $\Delta d/d_0=$-8\%, the valence band maximum at the Z point consists mainly of $d_{xz}$ and $d_{yz}$, or $d_{yz-xz}$ in terms of another orbital convention\cite{1998,2005}, and the conduction band minimum at the $\Gamma$ point mainly of $d_{x^2-y^2}$ orbital.

\subsection{Further discussions}

Furthermore, we show the strain dependence of the oxygen octahedron and the key bond lengths and bond angles in rutile and orthorhombic CrO$_2$ phases. In Fig. 6 we present the strain dependence of the bond length ($l_{b1}$) between Cr1 and O3 and the O-O bond length ($l_{\rm O}$) that can be defined as the distance between O1 and O2. We can see in Fig. 2 and Table II that when the biaxial compressive stress is applied, both $l_{b1}$ and $\alpha_2$ decrease nearly linearly, but $l_{b2}$ and $\alpha_1$ remain nearly unchanged. The O-O bond length $l_{\rm O}$ also decreases with the compressive stress increasing. It is clear that the O-O bond length $l_{\rm O}$ is compressed more easily than the horizontal bond length $l_{b1}$. This implies that it is more easy to compress the bond angle $\alpha_2$ than to compress the horizontal length $l_{b2}$. These changes of the bond lengths and angles mean that the oxygen octahedron is distorted and there is a small rotation around the z axis.

\begin{figure}[!htbp]
\centering  
\includegraphics[clip, width=7cm]{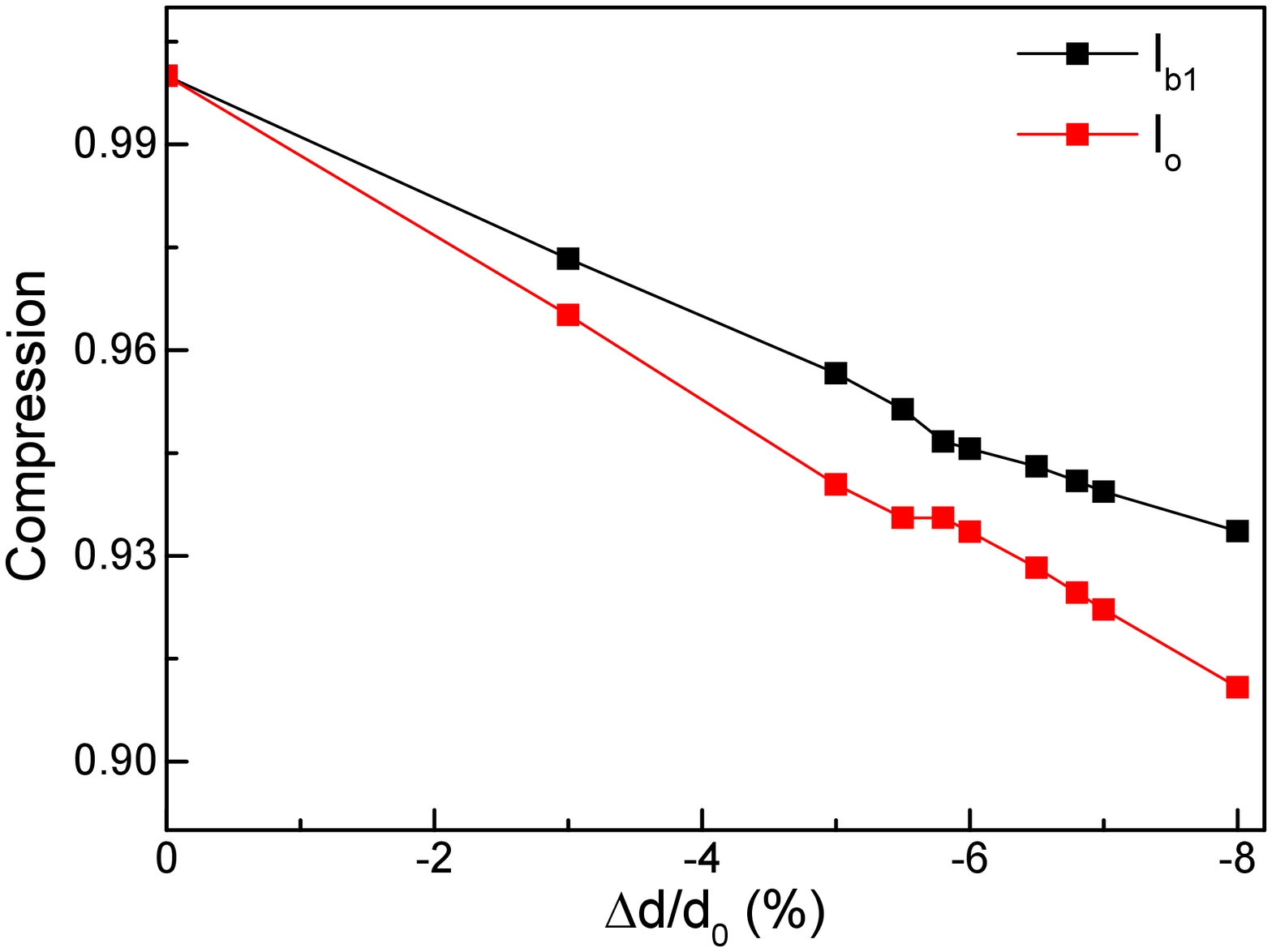}
\caption{The strain-dependent compression ratios of the horizontal bond ($l_{b1}$) and the O-O distance ($l_{\rm O}$). }\label{edge}
\end{figure}

\begin{figure}[!htbp]
\centering  
\includegraphics[clip, width=7.6cm]{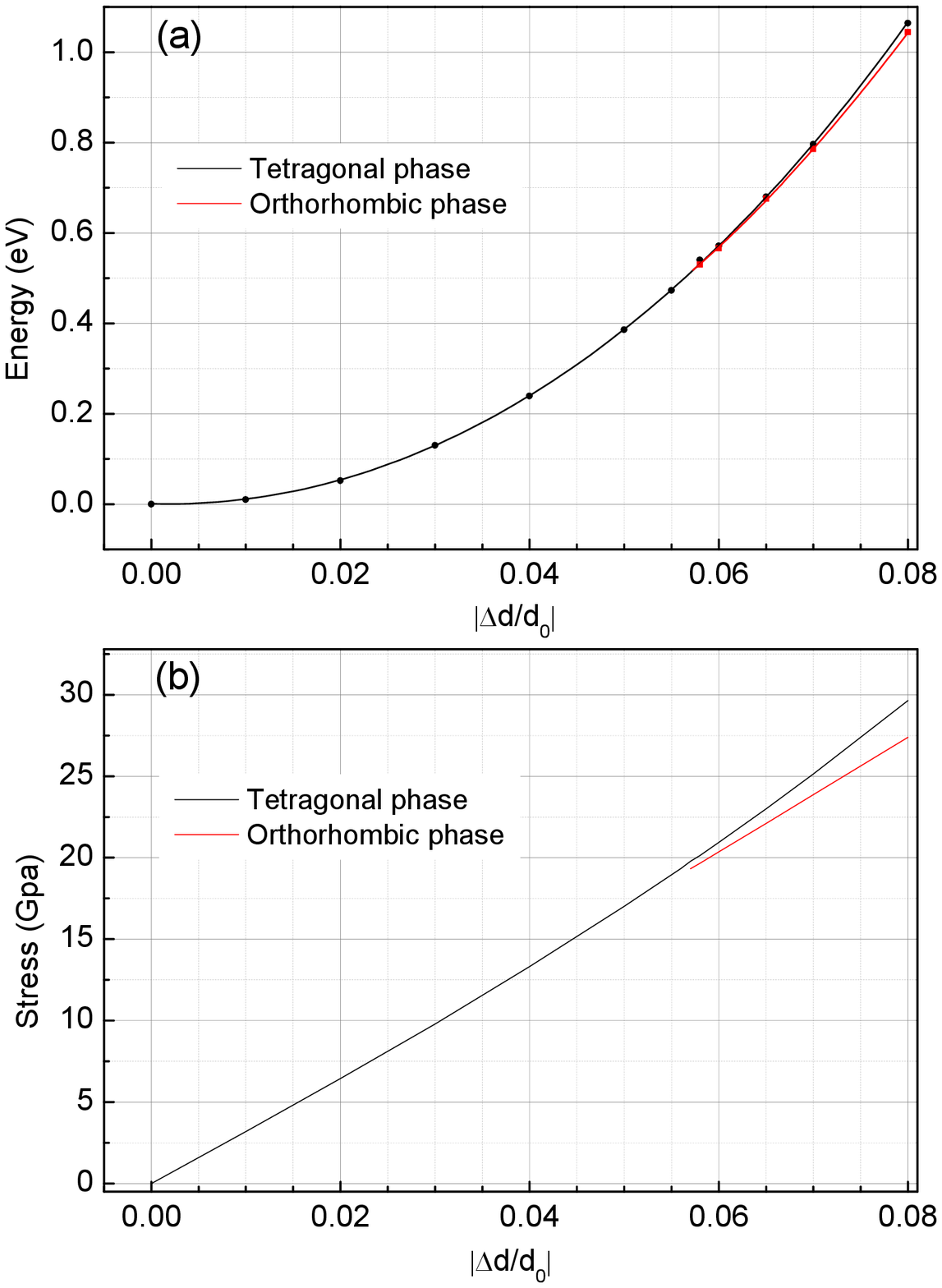}
\caption{The $|\Delta d/d_0|$ dependence of total energy (a) and stress (b). It can be seen that for $|\Delta d/d_0|\ge 5.6$\%, the total energy of the orthorhombic phase is only a little smaller than that of the tetragonal phase, but the stress is much smaller. }\label{edge}
\end{figure}

We present in Fig. 7 the elastic energy and stress as functions of the in-plane compressive ratio $\Delta d/d_0$. With $|\Delta d/d_0|\ge 5.6$\%, the rutile CrO$_2$ transits to the orthorhombic phase that features smaller energy and smaller stress for the same $|\Delta d/d_0|$. The stress can be realized by applying biaxial pressure, or growing the CrO$_2$ on substrates with smaller in-plane lattice constants.
We can realize such compression by growing the CrO$_2$ on good oxide substrates such as CoO, MgO, NiO, and BaSnO$_3$ \cite{add1,add2}. For these four substrates, the lattice mismatch is between -3.4\% and -7\%, and the stress is between 11.5 and 23.8 GPa. Accordingly, the first three heterostructures (CrO$_2$/CoO, CrO$_2$/MgO, CrO$_2$/NiO) should still be half-metallic, and the heterostructure CrO$_2$/BaSnO$_3$ should be semiconducting. It is interesting that EuO epitaxial thin films have been grown on SrTiO$_3$ \cite{euo-srtio3}, Si \cite{euo-si}, and MgO \cite{euo-mgo1,euo-mgo2}. Actually, a giant compressive strain of -22\% (due to lattice mismatch of 22\%) has been realized in crystalline epitaxial EuO/MgO(001) systems\cite{euo-mgo1,euo-mgo2}. Therefore, these strained CrO$_2$ thin films should be realizable.  Because the Curie temperature for such strain values should remain almost the same as that of equilibrium strain-free CrO$_2$, we can manipulate the transport properties of the CrO$_2$ at room temperature by growing it on appropriate substrates.

\section{Conclusion}

We have investigated the structural, magnetic, electronic properties of the rutile and orthorhombic CrO$_2$ phases under biaxial compressive stress through first-principles calculation. mBJ exchange potential is used to improve the description of the electronic structure. The biaxial compressive stress makes the in-plane lattice parameter $d=a$ (or $d=\sqrt{ab}$) decrease and the out-of-plane $c$ increase. Correspondingly, the two horizontal Cr-O and O-O bond-lengths and the side O-Cr-O bond-angle decrease linearly, but the oblique bond-lengths remain almost unchanged. There is a structural phase transition at $\Delta d/d_0$=-5.6\%. Our total energy comparison shows that the Curie temperature of the strained CrO$_2$ remains almost the same as that (near 400 K) without any stress. As for the electronic structure, the density of states near the Fermi level decreases with the $d$ becoming small and there is a half-metal-semiconductor transition at $\Delta d/d_0$=-6.1\%.
Full spin polarization can be realized for controllable concentration of carriers by usual gating or doping used in modern semiconductor technology. Our further analysis shows that this semiconducting ferromagnetic phase is driven by the biaxial-stress-induced distortion of the oxygen octahedron of Cr in the crystal structure. This biaxial compression can be realized by applying biaxial pressure or growing CrO$_2$ epitaxially on appropriate substrates. Therefore, these could pave a road to achieve good ferromagnetic semiconductors for spintronic applications.

\begin{acknowledgments}
This work is supported by the Nature Science Foundation of China (Grant No. 11574366), by the Strategic Priority Research Program of the Chinese Academy of Sciences (Grant No.XDB07000000), and by the Department of Science and Technology of China (Grant No. 2016YFA0300701). All the numerical calculations were performed in the Milky Way \#2 Supercomputer system at the National Supercomputer Center of Guangzhou, Guangzhou, China.
\end{acknowledgments}

\end{document}